\documentclass[aps,twocolumn]{revtex4}

\usepackage{graphics}
\usepackage{epsfig}

\usepackage{amsmath}

\newcommand{\bc}{\begin{center}}
\newcommand{\ec}{\end{center}}
\newcommand{\bt}{\begin{tabbing}}
\newcommand{\et}{\end{tabbing}} 
\newcommand{\be}{\begin{eqnarray*}}
\newcommand{\ee}{\end{eqnarray*}}
\newcommand{\bs}{\begin{slide}}
\newcommand{\es}{\end{slide}}

\begin{document}

\title{Motion in an Asymmetric Double Well}

\author{Alain J.~Brizard$^{1}$ and Melissa C.~Westland$^{2}$}
\affiliation{$^{1}$Department of Physics, Saint Michael's College, Colchester, VT 05439, USA \\ $^{2}$Department of Mathematics, Saint Michael's College, Colchester, VT 05439, USA}

\begin{abstract}
The problem of the periodic motion of a particle in an asymmetric double-well (quartic) potential is solved explicitly in terms of the Weierstrass and Jacobi elliptic functions. While the solution of the orbital motion is expressed simply in terms of the Weierstrass elliptic function, the period of oscillation is more directly expressed in terms of periods of the Jacobi elliptic functions. 
\end{abstract}

\begin{flushright}
July 3, 2016
\end{flushright}


\maketitle

\section{Introduction}

The double-well potential is an important paradigm in physics and chemistry. In classical mechanics, for example, it is used to study the motion of a particle either trapped in one of the two wells or moving with energy above the height of the barrier that separates the two wells \cite{Brizard_Lag}. For particles trapped in either of the two wells, the symmetry of the double-well potential implies that the oscillation periods are equal. When quantum tunneling through the barrier of a symmetric double-well potential is taken into account, however, the coupling of the solutions of the Schr\"{o}dinger equation for this two-state quantum problem \cite{Landau, Griffiths, SDW_1, SDW_2} leads to a splitting of the degenerate energy level. The ammonia-inversion problem is a well-known example of a symmetric double-well potential \cite{ammonia, Miller_1979}. Important quantum applications of a symmetric double-well potential include atom interferometry using Bose-Einstein condensates \cite{BEC_S1,BEC_S2}.

The addition of asymmetry in the double-well potential, where the two minima are no longer at the same energy level, introduces nontrivial modifications of the standard asymptotic treatment of quantum tunneling and energy-level splitting \cite{Miller_1979,ADW_1,ADW_2,ADW_3}. Quantum applications of an asymmetric double-well potential also provide a natural generalization of atom interferometry using Bose-Einstein condensates \cite{BEC_A1,BEC_A2,BEC_A3}.

The purpose of the present paper is to explore the classical orbits of a particle moving in an asymmetric double-well potential represented by a quartic polynomial. In particular, on an energy level that allows periodic 
motion in the shallow and deep wells, we prove that the oscillation periods for these orbits are equal. This result was recently derived by Levi \cite{Levi} based on the process of contour deformation on the 
Riemann sphere \cite{Ahlfors}. Here, we prove this result by deriving an explicit solution of the asymmetric double-well problem expressed in terms of the Weierstrass elliptic functions 
\cite{Brizard_elliptic,NIST_Weierstrass,Lawden,Brizard_2015}.

The remainder of this paper is organized as follows. In Sec.~\ref{sec:min_max}, we characterize an asymmetric quartic potential in terms of its two minima $(V_{c} \leq V_{a})$ and its single maximum $V_{b} \geq
 V_{a}$. Here, the asymmetric quartic potential is parameterized by a single asymmetry parameter $\delta$ that vanishes in the symmetric case. In Sec.~\ref{sec:energy}, the four turning points $(\xi_{1},\xi_{2},\xi_{3},
 \xi_{4})$ for the asymmetric quartic potential are expressed in terms of simple formulas parameterized by the energy value $E$ and the asymmetry parameter $\delta$. In Sec.~\ref{sec:Weierstrass}, solutions of the 
 motion in the asymmetric quartic potential are given in terms of the Weierstrass elliptic function by transforming the quartic energy equation into the standard Weierstrass cubic equation
\cite{NIST_Weierstrass,Lawden}. In Sec.~\ref{sec:Jacobi}, we express the solutions for the asymmetric quartic orbits in terms of the Jacobi elliptic functions \cite{NIST_Jacobi} with periods expressed in terms of the 
complete elliptic integral of the first kind. Lastly, in Sec.~\ref{sec:sum}, we briefly discuss a classical application of the present work when the asymmetry in the double-well potential is time dependent (e.g., by considering the undamped Duffing oscillator driven to chaotic behavior by a periodic force \cite{Holmes_1979,LL_chaos,Litak}).

\section{\label{sec:min_max}Local Minima and Maxima of an Asymmetric Double Well}

\begin{figure}
\epsfysize=2in
\epsfbox{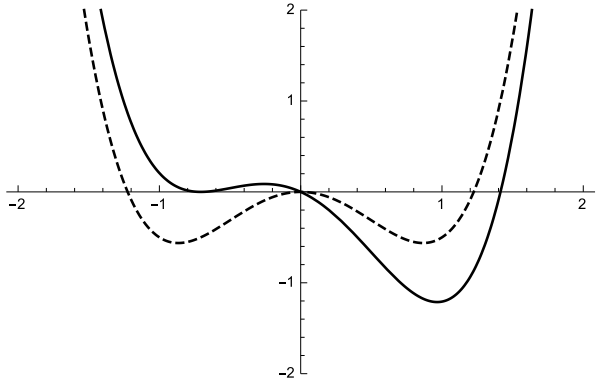}
\caption{Double-well potential \eqref{eq:double_well} for fixed values of $\delta \geq 0$: symmetric potential (dashed line) for $\delta(\varphi = \pi/2) = 0$ and asymmetric potential (solid line) for $\delta(\varphi = \pi/4) = 1/\sqrt{2}$.}
\label{fig:double_well}
\end{figure}

We consider an asymmetric double-well potential
\begin{equation}
V(x) \;\equiv\; x^{4} \;-\; \frac{3}{2}\,x^{2} \;-\; \delta\;x,
\label{eq:double_well}
\end{equation}
where the parameter $\delta$ represents the asymmetry in the potential. Figure \ref{fig:double_well} shows that if $\delta$ is in the range $|\delta| \leq 1$, the potential \eqref{eq:double_well} has two local minima 
$(x_{c} > x_{a})$ and one local maximum $x_{b}$ (with $x_{a} < x_{b} < x_{c}$), which are solutions of the cubic equation
\begin{equation}
V^{\prime}(x_{i}) \;=\; 4\;x_{i}^{3} \;-\; 3\,x_{i} \;-\; \delta \;=\; 0.
\label{eq:cubic_eq}
\end{equation}
The case $|\delta| > 1$, when a single local minimum remains, will not be considered in this work. The cubic equation \eqref{eq:cubic_eq} is a special case of the generic Weierstrass cubic equation $4\,x^{3} - 
g_{2}\,x - g_{3} = 0$ discussed in Appendix A. 

It is convenient to parameterize the asymmetry parameter in terms of the phase $0 \leq \varphi \leq \pi$:
\begin{equation}
\delta(\varphi) \;\equiv\; \cos\varphi \;=\; 4\,\cos^{3}\left(\frac{\varphi}{3}\right) \;-\; 3\,\cos\left(\frac{\varphi}{3}\right),
\label{eq:delta_def}
\end{equation}
so that the cubic roots of Eq.~\eqref{eq:cubic_eq} are thus expressed as
\begin{equation}
\left. \begin{array}{rcl}
x_{a}(\varphi) & = & -\;\cos[(\pi - \varphi)/3] \\
x_{b}(\varphi) & = & -\;\cos[(\pi + \varphi)/3] \\
x_{c}(\varphi) & = & \cos(\varphi/3)
\end{array} \right\},
\label{eq:V_prime_roots}
\end{equation}
and the roots \eqref{eq:V_prime_roots} satisfy the condition $x_{a} + x_{b} + x_{c} = 0$.

\begin{figure}
\epsfysize=3in
\epsfbox{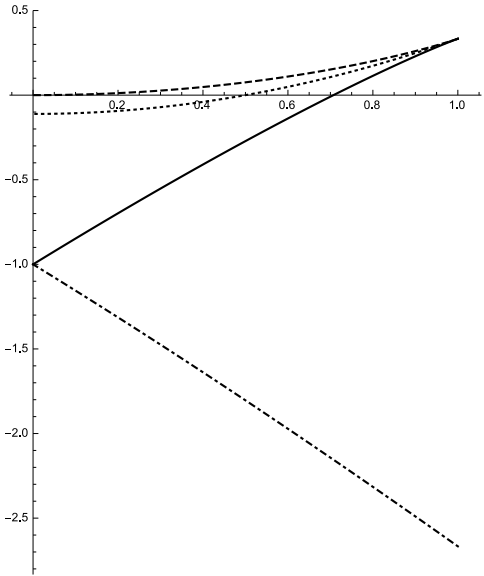}
\caption{Potential minima $\epsilon_{a}(\varphi) = 16 V_{a}(\varphi)/9$ (solid) and $\epsilon_{c}(\varphi) = 16 V_{c}(\varphi)/9$ (dot-dashed) and potential maximum $\epsilon_{b}(\varphi) = 16 V_{b}(\varphi)/9$ (dashed), defined in Eqs.~\eqref{eq:V_prime_roots}-\eqref{eq:min_max}, plotted as functions of 
$\delta(\varphi) = \cos\varphi$ from $\delta = 0$ to $\delta = 1$. The normalized energy level $\epsilon_{\delta}(\varphi) = (4\,\delta^{2} - 1)/9 \leq \epsilon_{b}(\varphi)$, defined in Eq.~\eqref{eq:epsilon_delta}, is also shown as a dotted curve. When $\delta = 1/\sqrt{2}$, we find 
$\epsilon_{c} < \epsilon_{a} = 0 < \epsilon_{\delta} = 1/9 < \epsilon_{b}$. }
\label{fig:epsilon_abc}
\end{figure}

The potential minima (at $x_{a} < x_{c}$) and maximum (at $x_{a} < x_{b} < x_{c}$) are
\begin{equation}
V_{i}(\varphi) \;=\; -\;\frac{3}{4}\,x_{i}(\varphi)\;\left( x_{i}(\varphi) \;+\frac{}{} \cos\varphi \right) \;\equiv\; \frac{9}{16}\,\epsilon_{i}(\varphi),
\label{eq:min_max}
\end{equation}
where we used $x_{i}^{3} = (3\,x_{i} + \delta)/4$ obtained from Eq.~\eqref{eq:cubic_eq}. Figure \ref{fig:epsilon_abc} shows the critical (normalized) energy levels 
\begin{equation}
-\,\frac{8}{3} \;\leq\; \epsilon_{c}(\varphi) \;\leq\; \epsilon_{a}(\varphi)  \;\leq\; \epsilon_{b}(\varphi) \;\leq\; \frac{1}{3} 
\label{eq:epsilon_i}
\end{equation}
as functions of the asymmetry parameter $\delta(\varphi) = \cos\varphi$ from $\delta = 0$ to $\delta = 1$.

For the numerical case to be studied later in this paper, we choose $\varphi = \pi/4$ so that $\delta = 1/\sqrt{2}$ (see Fig.~\ref{fig:epsilon_abc}) and the potential extrema \eqref{eq:min_max} are
\begin{equation}
\left. \begin{array}{rcl}
\epsilon_{a} & = & 0 \\
 &  & \\
\epsilon_{b} & = & \frac{2}{3}\,\sqrt{3} - 1 \;=\; 0.1547... \\
 &  & \\
 \epsilon_{c} & = & -\,\frac{2}{3}\,\sqrt{3} - 1  \;=\; -\,2.1547...
\end{array} \right\}.
\label{eq:epsilon_numerical}
\end{equation}
These values will be used as markers as we derive explicit solutions for the periodic motion of a particle moving in the asymmetric double-well potential 
\eqref{eq:double_well}, as shown in Figs.~\ref{fig:chi}-\ref{fig:Theta}.

\section{\label{sec:energy}Energy Levels in an Asymmetric Double Well}

\begin{figure}
\epsfysize=2.5in
\epsfbox{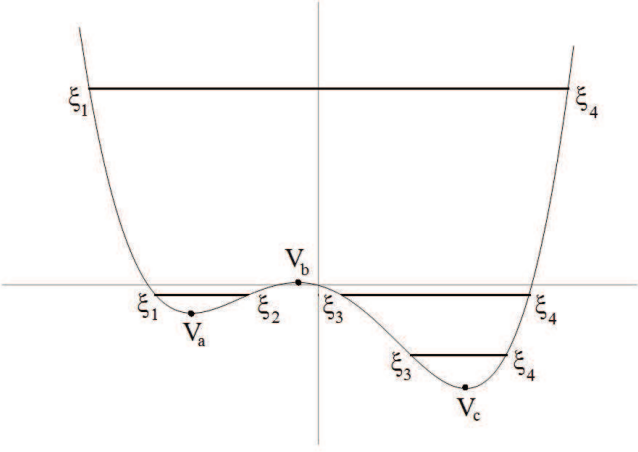}
\caption{Turning points $\xi_{k}$ $(k = 1,2,3,4)$ for an asymmetric double well with $V_{c} < V_{a} < V_{b}$. The turning points merge at $\xi_{3} = \xi_{4} = x_{c}$ ($E = V_{c}$), $\xi_{1} = \xi_{2} = x_{a}$ ($E = V_{a}$), and $\xi_{2} = \xi_{3} = x_{b}$ ($E = V_{b}$). When $E > V_{b}$, the points $\xi_{1} < \xi_{4}$ are real and $\xi_{2} = \xi_{3}^{*}$.}
\label{fig:double_well_2}
\end{figure}

All orbits in the potential \eqref{eq:double_well} are bounded orbits with either two real turning points or four real turning points. We now study the turning points $\xi_{k}$ ($k = 1,2,3,4$) associated with an energy level $E \equiv 9\,\epsilon/16$, which are roots of the quartic energy equation
\begin{equation}
\xi_{k}^{4} \;-\; \frac{3}{2}\,\xi_{k}^{2} \;-\; \delta\;\xi_{k} \;-\; \frac{9\,\epsilon}{16} \;=\; 0.
\label{eq:energy_quartic}
\end{equation}
These turning points verify the identities
\begin{equation}
\left. \begin{array}{rcl}
\sum_{i}\xi_{i} & = & 0 \\
\sum_{i< j}\xi_{i}\,\xi_{j}  & = & -\,3/2 \\
\sum_{i< j < k}\xi_{i}\,\xi_{j}\,\xi_{k} & = & \delta \\
\xi_{1}\xi_{2}\xi_{3}\xi_{4} & = & -9\,\epsilon/16
\end{array} \right\}.
\label{eq:turning_identities}
\end{equation} 
Table \ref{tab:energy_levels} summarizes the energy-levels for $0 < \delta < 1$ (with $V_{c} < V_{a} < V_{b}$) and the possible turning-point scenarios. For the case $V_{c} < E < V_{a}$ (i.e., the energy level is located between the two local minima), only two real turning points exist ($\xi_{3} < \xi_{4}$) and periodic motion is located in the deep potential well on the right side of Fig.~\ref{fig:double_well_2}. For the case $V_{a} < E < V_{b}$ (i.e., the energy level is located between the highest local minimum and the local maximum), all four turning points are real ($\xi_{1} < \xi_{2} < \xi_{3} < \xi_{4}$) and periodic motion in possible in either the shallow potential well (left side) or the deep potential well (right side). Lastly, for $E > V_{b}$ (i.e., the energy level is located above the local maximum), only two real roots exist ($\xi_{1} < \xi_{4}$) and the motion is periodic above the two wells.

\begin{table}
\caption{\label{tab:energy_levels}Turning Points for $0 < \delta < 1$.} 
\begin{ruledtabular} 
\begin{tabular}{lcc}
Energy                         & Real                                                          & Complex    \\ 
Levels                          & Turning Points                                            & Turning Points \\ \hline 
$E > V_{b}$                 & $\xi_{1} < \xi_{4}$                                      & $\xi_{2} = \xi_{3}^{*}$ \\
$E = V_{b}$                 &  $\xi_{1} < \xi_{2} = \xi_{3} < \xi_{4}$         & \\
$V_{a} < E < V_{b}$    & $\xi_{1} < \xi_{2} < \xi_{3} < \xi_{4}$          &                          \\ 
$E = V_{a}$                 &  $\xi_{1} = \xi_{2} < \xi_{3} < \xi_{4}$         & \\
$V_{c} \leq E < V_{a}$    & $\xi_{3} \leq \xi_{4}$                                       & $\xi_{1} = \xi_{2}^{*}$
\end{tabular}
\end{ruledtabular}
\end{table}

\subsection{Cubic root $\chi(\mu,\nu)$}

\begin{figure}
\epsfysize=2in
\epsfbox{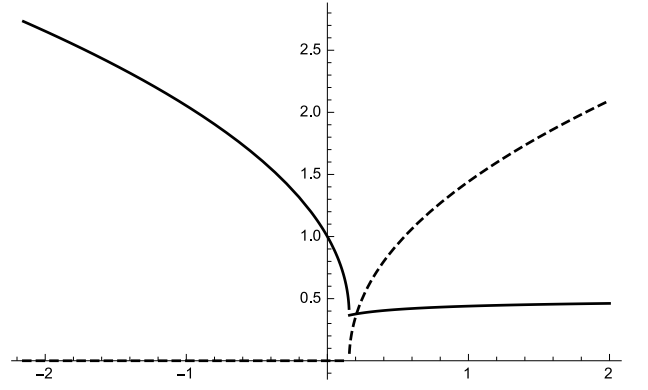}
\caption{Plots of the real (solid) and imaginary (dashed) parts of $\chi(\mu,\nu)$ as functions of $\epsilon$ for the case 
$\delta(\varphi = \pi/4) = 1/\sqrt{2}$. Here, $\chi(\mu,\nu)$ becomes complex-valued for $\epsilon > \epsilon_{b}$.}
\label{fig:chi}
\end{figure}

\begin{figure}
\epsfysize=2in
\epsfbox{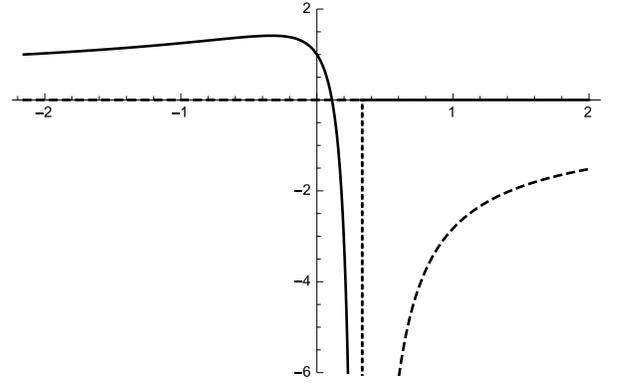}
\caption{Plots of the real (solid) and imaginary (dashed) parts of $\eta(\mu,\nu)$ as functions of $\epsilon$ for the case 
$\delta(\varphi = \pi/4) = 1/\sqrt{2}$. Here, $\eta_{a} = 1 = \eta_{c} = 1$, so that $\eta \geq 1$ for $\epsilon_{c} \leq \epsilon \leq \epsilon_{a}$, and $\eta_{b} = -1$, so that $-1 \leq \eta \leq 1$ for $\epsilon_{a} \leq \epsilon \leq \epsilon_{b}$. Lastly, $\eta \rightarrow -\,\infty$ for $\epsilon \rightarrow \frac{1}{3}^{-}$ (dotted line), and 
$\eta = -i\,|\eta|$ for $\epsilon > \frac{1}{3}$.}
\label{fig:cos_psi}
\end{figure}

In order to obtain explicit and compact expressions for the turning points $\xi_{k}(\epsilon,\delta)$ ($k = 1,2,3,4$), we define the parametric functions 
\begin{equation}
\left. \begin{array}{rcl}
\nu(\epsilon) & \equiv & 1 \;-\; 3\,\epsilon \\
 &  & \\
\mu(\epsilon,\delta) & \equiv & 4\,\delta^{2} - (1 + 9\,\epsilon)
\end{array} \right\},
\label{eq:nu_mu}
\end{equation}
and introduce the function 
\begin{eqnarray}
\chi(\mu,\nu) & \equiv & \frac{1}{2} \left(\mu - \sqrt{\mu^{2} \;-\; \nu^{3}}\right)^{1/3} \nonumber \\
 &  &+\; \frac{\nu}{2}\;\left(\mu - \sqrt{\mu^{2} \;-\; \nu^{3}}\right)^{-1/3},
 \label{eq:chi_def}
\end{eqnarray} 
which is a solution of the cubic equation
\begin{equation}
4\,\chi^{3} \;-\; 3\,\nu\;\chi \;-\; \mu \;=\; 0.
\label{eq:chi_cubic}
\end{equation}
Figure \ref{fig:chi} shows that $\chi(\mu,\nu)$ is real in the energy range $\epsilon_{c} \leq \epsilon \leq \epsilon_{b} < 1/3$, so that Eq.~\eqref{eq:chi_def} may be expressed as
\begin{equation}
\chi(\mu,\nu) \;\equiv\; \nu^{1/2}\;\cos\left(\frac{1}{3}\,\psi(\eta)\right)
\label{eq:chi_cb}
\end{equation}
where $\nu > 0$ in that range and
\begin{equation} 
\cos\psi(\eta) \;=\; \mu/\nu^{3/2} \equiv \eta(\mu,\nu).
\label{eq:psi_def}
\end{equation}
In Eq.~\eqref{eq:psi_def}, the  function $\eta(\mu,\nu)$ is not restricted to the interval $-1 \leq \eta \leq 1$ (see Fig.~\ref{fig:cos_psi}), and, hence,  the phase $\psi(\eta)$ may be complex valued. Based on Fig.~\ref{fig:chi} and the solution \eqref{eq:chi_cb}, the phase $\psi(\eta)$ is, therefore, defined as
\begin{equation}
\psi(\eta) \;=\; \left\{ \begin{array}{lcr}
i\,\cosh^{-1}(\eta) & \hspace*{0.1in} & (\epsilon_{c} \leq \epsilon \leq \epsilon_{a}) \\
\cos^{-1}(\eta) & &  (\epsilon_{a} \leq \epsilon \leq \epsilon_{b}) \\
\pi - i\, \cosh^{-1}(|\eta|) &  & (\epsilon_{b} \leq \epsilon < 1/3) \\
\frac{\pi}{2} + i\, \sinh^{-1}(|\eta|) & & (\epsilon > 1/3)
\end{array} \right.
\label{eq:psi_path}
\end{equation}
and, when $\epsilon > 1/3$ (i.e., $\nu < 0$), we use $\nu^{1/2} = i\,|\nu|^{1/2}$ in Eq.~\eqref{eq:chi_cb}, with $\mu = -\,|\mu|$ and $\eta = -i\,|\eta|$ in Eq.~\eqref{eq:psi_def}.

At the minima $\epsilon_{c} < \epsilon_{a}$ and the local maximum $\epsilon_{b} > \epsilon_{a}$, we find (for $i = a, b, c$)
\begin{equation}
\left. \begin{array}{rcl}
\nu_{i} & = & 1 \;-\; 3\,\epsilon_{i} \;\equiv\; (4\,x_{i}^{2} - 1)^{2} \\
 &  & \\
 \mu_{i} & = & 4\,\delta^{2} - (1 \;+\; 9\,\epsilon_{i}) \;\equiv\; (4\,x_{i}^{2} - 1)^{3}
\end{array} \right\},
\label{eq:numu_i}
\end{equation}
so that $\eta_{a} = \eta_{c} = 1$ and $\eta_{b} = -1$. Hence, we obtain $\psi_{a} = 0 = \psi_{c}$ and $\psi_{b} = \pi$, and Eq.~\eqref{eq:chi_cb} yields
\begin{equation}
\left. \begin{array}{l}
\chi_{a}(\varphi) \;=\; 4\,x_{a}(\varphi)^{2} \;-\; 1 \\
\chi_{b}(\varphi) \;=\; \frac{1}{2} \;-\; 2\,x_{b}(\varphi)^{2} \\
\chi_{c}(\varphi) \;=\; 4\,x_{c}(\varphi)^{2} \;-\; 1
\end{array} \right\}.
\label{eq:chi_i}
\end{equation}
We note that $\eta(\mu,\nu)$ vanishes when $\mu = 0$ at
\begin{equation}
\epsilon \;=\; \epsilon_{\delta} \,\equiv\, \frac{1}{9}\,(4\,\delta^{2} - 1), 
\label{eq:epsilon_delta}
\end{equation}
where $-1/9 \leq \epsilon_{\delta} \leq \epsilon_{b}$ (see the dotted curve in Fig.~\ref{fig:epsilon_abc}), so that $\psi_{\delta} \equiv \pi/2$, $\nu_{\delta} = (4/3)\,\sin^{2}\varphi$, and $\chi_{\delta} = \sin\varphi$. 

\subsection{Quartic turning points}

Next, using the definition $\chi \equiv 4\,\sigma^{2} - 1$, the four roots of the quartic energy equation \eqref{eq:energy_quartic} are expressed as
\begin{eqnarray}
\xi_{1}(\sigma,\delta) & \equiv & -\;\sigma \;-\; \frac{1}{2}\;\sqrt{3 - 4\,\sigma^{2} - \delta/\sigma},
\label{eq:xi_1} \\
\xi_{2}(\sigma,\delta) & \equiv & -\;\sigma \;+\; \frac{1}{2}\;\sqrt{3 - 4\,\sigma^{2} - \delta/\sigma},
\label{eq:xi_2} \\
\xi_{3}(\sigma,\delta) & \equiv & \sigma \;-\; \frac{1}{2}\;\sqrt{3 - 4\,\sigma^{2} + \delta/\sigma},
\label{eq:xi_3} \\
\xi_{4}(\sigma,\delta) & \equiv & \sigma \;+\; \frac{1}{2}\;\sqrt{3 - 4\,\sigma^{2} + \delta/\sigma},
\label{eq:xi_4}
\end{eqnarray}
where the radicands
\[ 3 - 4\,\sigma^{2} \mp \delta/\sigma \;=\; \left\{ \begin{array}{l}
V^{\prime}(-\sigma)/\sigma \\
 \\
 -\,V^{\prime}(\sigma)/\sigma 
 \end{array} \right. \]
 are expressed in terms of the potential derivative \eqref{eq:cubic_eq}. The roots \eqref{eq:xi_1}-\eqref{eq:xi_4} satisfy the identities \eqref{eq:turning_identities}, with
\begin{eqnarray}
\xi_{1}\xi_{2}\xi_{3}\xi_{4} & = & 4\,\sigma^{4} \;-\; 3\,\sigma^{2} \;+\; \frac{1}{16} \left(9 - \frac{\delta^{2}}{\sigma^{2}}\right) \nonumber \\
 & = & \frac{1}{64\,\sigma^{2}}\left( 4\,\chi^{3} \;-\frac{}{} 3\,\chi \;-\; (\mu + 9\,\epsilon)\right) \nonumber \\
  & = & -\;\frac{9\epsilon\,(\chi + 1)}{64\,\sigma^{2}} =  -\;\frac{9\,\epsilon}{16},
\end{eqnarray}
which follows from Eq.~\eqref{eq:chi_cubic} and the definitions \eqref{eq:nu_mu}. The two roots \eqref{eq:xi_1}-\eqref{eq:xi_2} merge $\xi_{1} = \xi_{2} = x_{a}$ at $\sigma = -\,x_{a}$, where $V^{\prime}(x_{a}) = 0$. Similarly, the two roots  \eqref{eq:xi_3}-\eqref{eq:xi_4} merge $\xi_{3} = \xi_{4} = x_{b}$ at $\sigma = x_{b}$, where $V^{\prime}(x_{b}) = 0$.

In numerical applications, we first select an asymmetry parameter $\delta(\varphi) = \cos\varphi$ and then choose an energy parameter $\epsilon = 16\,E/9$ in one of the five ranges 
\begin{equation}
\left. \begin{array}{lcr}
{\rm (I)} & \hspace*{0.5in} &  \epsilon_{c}(\varphi) < \epsilon < \epsilon_{a}(\varphi) \\
{\rm (II.a)} & & \epsilon_{a}(\varphi) < \epsilon < \epsilon_{\delta}(\varphi) \\
{\rm (II.b)} & & \epsilon_{\delta}(\varphi) < \epsilon <  \epsilon_{b}(\varphi) \\
{\rm (III)} & & \epsilon_{b}(\varphi) < \epsilon < 1/3 \\
{\rm (IV)} & & \epsilon > 1/3
\end{array} \right\}. 
\label{eq:epsilon_region}
\end{equation}
For each pair $(\epsilon,\delta)$, we therefore obtain a pair $(\mu,\nu)$ defined in Eq.~\eqref{eq:nu_mu}, from which we find $\chi(\mu,\nu)$ from 
Eq.~\eqref{eq:chi_cb} and $\sigma(\mu,\nu) = \frac{1}{2}\,\sqrt{\chi(\mu,\nu)+1}$, from which we calculate the four turning points \eqref{eq:xi_1}-\eqref{eq:xi_4}. Using one of the turning points as an initial condition, we then look for a solution for the orbit $x(t)$.

\section{\label{sec:Weierstrass}Weierstrass Solutions}

The motion of a particle (of unit mass) in the potential \eqref{eq:double_well} is expressed as a solution to the ordinary differential equation
\begin{eqnarray}
\frac{1}{2}\;\left(\frac{dx}{dt}\right)^{2} & = & \frac{9\,\epsilon }{16} \;-\; \left( x^{4} \;-\; \frac{3}{2}\,x^{2} \;-\; \delta\;x \right) 
\label{eq:motion_double_x} \\
 & \equiv & -\;(x - \xi_{1})\,(x - \xi_{2})\,(x - \xi_{3})\,(x - \xi_{4}) \nonumber \\
  & = & (\xi_{4} - x)\,(x - \xi_{3})\,(x - \xi_{2})\,(x - \xi_{1}),
\nonumber
\end{eqnarray}
where we used the fact that $\xi_{1} \leq x \leq \xi_{4}$ for all orbits.We now investigate orbits in the quartic potential \eqref{eq:double_well} for the initial condition $x(0) = \xi_{1}$ (for $E > V_{a}$) or $x(0) = \xi_{4}$ (for $E > V_{c}$). 

Since the right side of Eq.~\eqref{eq:motion_double_x} involves a quartic polynomial in $x$, we could seek a solution for $x(t)$ in terms of the Jacobi elliptic functions. Because the quartic polynomial is not symmetric, however, we seek a solution in terms of the Weierstrass elliptic function by transforming 
Eq.~\eqref{eq:motion_double_x} into a new differential equation involving a cubic polynomial with the substitution $x(t) = \xi_{0} \pm 1/s(t)$, where the initial condition $x(0) = \xi_{0}$ is chosen as one of the four turning points \eqref{eq:xi_1}-\eqref{eq:xi_4} and the new variable $s(t)$ satisfies the initial condition $s(0) = \infty$.

\subsection{Periodic motion}

Before deriving the Weierstrass solutions for Eq.~\eqref{eq:motion_double_x}, we derive explicit expressions for the periods $T_{12}$ and $T_{34}$. When all turning points $\xi_{1} \leq \xi_{2} \leq \xi_{3} < \xi_{4}$ are real (i.e., for $\epsilon_{a} \leq \epsilon \leq \epsilon_{b}$), the two periods $T_{12}$ and $T_{34}$ for bounded periodic motions between $\xi_{1} < x < \xi_{2}$ and $\xi_{3} < x < \xi_{4}$ are defined from Eq.~\eqref{eq:motion_double_x}, 
respectively, as
\begin{eqnarray}
T_{12} & = & \int_{\xi_{1}}^{\xi_{2}}\frac{\sqrt{2}\,dx}{\sqrt{(\xi_{4} - x)(\xi_{3}-x)\,(\xi_{2}-x)(x - \xi_{1})}} \label{eq:T_12} \\
 & \equiv & \oint_{C_{12}}\frac{dz}{\sqrt{2\,(\xi_{4} - z)(\xi_{3}-z)\,(\xi_{2}-z)(z - \xi_{1})}}, \nonumber \\
T_{34} & = & \int_{\xi_{3}}^{\xi_{4}}\frac{\sqrt{2}\,dx}{\sqrt{(\xi_{4} - x)(x-\xi_{3})\,(x-\xi_{2})(x - \xi_{1})}} \label{eq:T_34} \\
 & \equiv & \oint_{C_{34}}\frac{dz}{\sqrt{2\,(\xi_{4} - z)(z-\xi_{3})\,(z-\xi_{2})(z - \xi_{1})}}. \nonumber
\end{eqnarray}
Here, the contours $C_{12}$ and $C_{34}$ are drawn in the complex $z$-plane around the respective branch cuts from $\xi_{1}$ to $\xi_{2}$ and $\xi_{3}$ to $\xi_{4}$.  When viewed on the Riemann sphere \cite{Ahlfors} (i.e., the extended complex plane defined as the complex plane plus the point at infinity), however, the contour $C_{12}$ can be continuously deformed into the contour $C_{34}$ (since the integrand has no singularity at infinity) and, therefore, the two periods are equal \cite{Levi}
\begin{equation}
T_{12}(\epsilon,\delta) \;\equiv\; T_{34}(\epsilon,\delta),
\label{eq:T_12=34}
\end{equation} 
for any pair $(\epsilon,\delta)$. Hence, the period of motion in one well can be computed by calculating the period of motion in the other well, including when the normalized energy 
$\epsilon$ is equal to a minimum value (i.e., $\epsilon = \epsilon_{a}$ or $\epsilon_{c}$).

When the energy $\epsilon = \epsilon_{a}$ (corresponding to the shallow-well minimum), we find 
\begin{equation}
\left. \begin{array}{rcl}
\xi_{1} & = & \xi_{2} \;=\; -\,\sigma \equiv x_{a} \\
\xi_{4} & = & \sigma \;+\; (3/2 - 2 \sigma^{2})^{1/2} \\
\xi_{3} & = & \sigma - (3/2 - 2 \sigma^{2})^{1/2}
\end{array} \right\},
\end{equation}
so that the period \eqref{eq:T_34} becomes
\begin{eqnarray}
T_{34}(\epsilon_{a}) & = & \oint_{C_{34}}\frac{dz}{\sqrt{2\,(\xi_{4} - z)(z-\xi_{3})\,(z+\sigma)^{2}}} \nonumber \\
 & = & \oint_{C_{a}}\frac{dz}{\sqrt{2\,(\xi_{4} - z)(z-\xi_{3})\,(z+\sigma)^{2}}} \nonumber \\
  & = & \frac{2\pi\,i}{\sqrt{2\,(\xi_{4}+\sigma)(-\sigma-\xi_{3})}}  \nonumber \\
  & = & \frac{2\pi}{\sqrt{8\,\sigma^{2} - (3 - 4 \sigma^{2})}} \;=\; \frac{2\pi}{\sqrt{3\,(4\sigma^{2} - 1)}} \nonumber \\
   & = & \frac{2\pi}{\sqrt{3\,(4\,x_{a}^{2}-1)}}  \;=\; \frac{2\pi}{\sqrt{V^{\prime\prime}(x_{a})}},
 \label{eq:T34_a}
  \end{eqnarray}
where the contour $C_{34}$ was continuously deformed to the contour $C_{a}$ around the single pole at $z = -\sigma = x_{a}$ and the residue theorem \cite{Ahlfors} has been used. The identity \eqref{eq:T_12=34} then yields the standard result \cite{Brizard_Lag}
\begin{equation}
T_{a} \;\equiv\; T_{34}(\epsilon_{a}) \;=\; \frac{2\pi}{\sqrt{V^{\prime\prime}(x_{a})}}.
\label{eq:T_a}
\end{equation}
Similarly, when the energy $\epsilon = \epsilon_{c}$ (corresponding to the deep-well minimum), we find 
\begin{equation}
\left. \begin{array}{rcl}
\xi_{3} & = & \xi_{4} \;=\; \sigma \equiv x_{c} \\
\xi_{1} & = & -\;\sigma \;-\; (3/2 - 2 \sigma^{2})^{1/2} \\
\xi_{2} & = & -\;\sigma \;+\; (3/2 - 2 \sigma^{2})^{1/2}
\end{array} \right\},
\end{equation}
so that the period \eqref{eq:T_12} becomes
\begin{eqnarray}
T_{12}(\epsilon_{c}) & = & \oint_{C_{12}}\frac{dz}{\sqrt{2\,(z - \sigma)^{2}\,(\xi_{2}-z)(z - \xi_{1})}} \nonumber \\
 & = & \oint_{C_{c}}\frac{dz}{\sqrt{2\,(z - \sigma)^{2}\,(\xi_{2}-z)(z - \xi_{1})}} \nonumber \\
 & = & \frac{2\pi\,i}{\sqrt{2\,(\xi_{2}-\sigma)(\sigma-\xi_{1})}}  \nonumber \\
   & = & \frac{2\pi}{\sqrt{8\,\sigma^{2} - (3 - 4 \sigma^{2})}} \;=\; \frac{2\pi}{\sqrt{3\,(4\sigma^{2} - 1)}} \nonumber \\
   & = & \frac{2\pi}{\sqrt{3\,(4\,x_{c}^{2}-1)}}  \;=\; \frac{2\pi}{\sqrt{V^{\prime\prime}(x_{c})}},
 \label{eq:T12_c}
\end{eqnarray}
which again yields the standard result 
\begin{equation}
T_{c} \;\equiv\; T_{12}(\epsilon_{c}) \;=\; \frac{2\pi}{\sqrt{V^{\prime\prime}(x_{c})}}.
\label{eq:T_c}
\end{equation}
Equations \eqref{eq:T_a} and \eqref{eq:T_c} can easily be obtained from the equation of motion $\ddot{x} = -\,V^{\prime}(x)$ by linearizing it about a stable equilibrium point $x_{0} \equiv x - \delta x$, where $V^{\prime}(x_{0}) = 0$ and $V^{\prime\prime}(x_{0}) > 0$, which yields the linearized equation $\delta\ddot{x} = -\,V^{\prime\prime}(x_{0})\;\delta x$, with a periodic sinusoidal solution with period $T_{0} = 2\pi/\sqrt{V^{\prime\prime}(x_{0})}$.

\begin{table}
\caption{\label{tab:omega_asymm}Period of oscillation for the asymmetric double well for the five ranges defined in Eq.~\eqref{eq:epsilon_region}.}
\begin{ruledtabular} 
\begin{tabular}{rrcc}
            &                                                                         & $(g_{2},g_{3},\Delta)$    & $T(\epsilon)$                                                              \\ \hline
            &  $\epsilon = \epsilon_{c}$                                & $(+,+,0)$                        & $2\,\Omega_{0}(\epsilon_{c})$                                      \\
(I)         & $\epsilon_{c} < \epsilon < \epsilon_{a}$          & $(+,+,-)$                        & $2\,\Omega$                                                               \\
            & $\epsilon = \epsilon_{a}$                                 & $(+,+,0)$                       & $2\,\Omega_{0}(\epsilon_{a})$                                       \\
(II.a)        & $\epsilon_{a} < \epsilon < \epsilon_{\delta}$   & $(+,+,+)$                       & $2\,\omega$                                                                \\
             & $\epsilon = \epsilon_{\delta}$                         & $(+,0,+)$                       & $2\,\omega_{0}$                  \\
(II.b)       & $\epsilon_{\delta} < \epsilon < \epsilon_{b}$   & $(+,-,+)$                        & $2\,|\omega^{\prime}|$                                                \\
             & $\epsilon = \epsilon_{b}$                                & $(+,-,0)$                        & $\infty$                                                                         \\
 (III)      & $\epsilon_{b} < \epsilon < \frac{1}{3}$            & $(+,-,-)$                         & $2\,|\Omega^{\prime}|$                                                \\
             & $\epsilon = \frac{1}{3}$                                   & $(0,-,-)$                         & $2\,{\rm Re}(\omega_{1}^{\rm e})$ \\
(IV)        &$\epsilon > \frac{1}{3}$                                    & $(-,-,-)$                          & $2\,{\rm Re}(\omega_{1})$                                           \\ 
\end{tabular}
\end{ruledtabular}
\end{table}

The periods of oscillation for the asymmetric double well for the five ranges defined in Eq.~\eqref{eq:epsilon_region} are summarized in Table \ref{tab:omega_asymm}. The calculations of these periods are based on the Weierstrass half-periods \cite{Brizard_2015}
\begin{eqnarray}
\omega_{1}^{\pm}(g_{2},g_{3}) & = & \int_{e_{1}}^{\infty} \frac{ds}{\sqrt{4\,s^{3} - g_{2}\,s - g_{3}}}, \label{eq:omega_1} \\
\omega_{3}^{\pm}(g_{2},g_{3}) & = & \pm\,i\;\int_{-\infty}^{e_{3}} \frac{ds}{\sqrt{|4\,s^{3} - g_{2}\,s - g_{3}|}}, \label{eq:omega_3}
\end{eqnarray}
where 
\begin{equation}
\left. \begin{array}{rcl}
g_{2} & = & (3/4)\,\nu \\
g_{3} & = & \mu/8 \\
\Delta & = & (27/64)\,(\nu^{3} - \mu^{2}) 
\end{array} \right\},
\end{equation}
and the cubic roots are
\begin{equation}
\left. \begin{array}{rcl}
e_{1}(\mu,\nu) & = & \frac{1}{2}\,\sqrt{\nu}\;\cos\left(\frac{1}{3}\,\psi(\eta)\right) \\
 &  & \\
e_{2}(\mu,\nu) & = & -\;\frac{1}{2}\,\sqrt{\nu}\;\cos\left(\frac{\pi}{3} + \frac{1}{3}\,\psi(\eta)\right)  \\
  &  & \\
e_{3}(\mu,\nu) & = & -\;\frac{1}{2}\,\sqrt{\nu}\;\cos\left(\frac{\pi}{3} - \frac{1}{3}\,\psi(\eta)\right) 
\end{array} \right\},
\label{eq:root_period}
\end{equation} 
with the phase $\psi(\eta)$ defined in Eq.~\eqref{eq:psi_path}. When $(g_{2},g_{3}) = (+,+)$ (i.e., $\epsilon_{c} < \epsilon < \epsilon_{\delta}$), we find the real half-periods \cite{Brizard_2015}
\begin{equation}
\omega_{1}^{+} \;=\; \left\{ \begin{array}{lcr}
\omega &  & (\Delta > 0) \\
 &  & \\
 \Omega &  & (\Delta < 0)
 \end{array} \right.
 \end{equation}
 and the complex half-periods
 \begin{equation}
\omega_{3}^{+} \;=\; \left\{ \begin{array}{lcr}
\omega^{\prime} = i\,|\omega^{\prime}| &  & (\Delta > 0) \\
 &  & \\
 i\,|\Omega^{\prime}| - \Omega/2 &  & (\Delta < 0)
 \end{array} \right.
 \end{equation}
 while, for $(g_{2},g_{3}) = (+,-)$ (i.e., $\epsilon_{\delta} < \epsilon < \frac{1}{3}$), we find \cite{Brizard_2015}
 \begin{equation}
  \omega_{1}^{-} \;=\; -i\,\omega_{3}^{+} \;=\; \left\{ \begin{array}{lcr}
|\omega^{\prime}| &  & (\Delta > 0) \\
 &  & \\
|\Omega^{\prime}| + i\,\Omega/2 &  & (\Delta < 0)
 \end{array} \right.
 \end{equation}
 with $e_{1}^{-} = -\,e_{3}^{+}$ and $e_{3}^{-} = -\,e_{1}^{+}$ used in Eqs.~\eqref{eq:omega_1}-\eqref{eq:omega_3}. The periods in Table \ref{tab:omega_asymm} are all given in terms of $2\,{\rm Re}(\omega_{1})$.
 The special cases associated with $\Delta = 0$ (i.e., for the critical energies $\epsilon = \epsilon_{c}$, $\epsilon_{a}$, or $\epsilon_{b}$) yield the real half-periods
 \begin{equation}
 \left. \begin{array}{rcl}
 \omega_{1}^{+}(\epsilon_{c}) & = & \Omega_{0}(\epsilon_{c}) \;\equiv\; \pi/\sqrt{3\,\chi_{c}} \\
 \omega_{1}^{+}(\epsilon_{a}) & = & \Omega_{0}(\epsilon_{a}) \;\equiv\; \pi/\sqrt{3\,\chi_{a}} \\
 \omega_{1}^{-}(\epsilon_{b}) & = & \infty
 \end{array} \right\}.
 \end{equation}
 
 In Sec.~\ref{sec:Jacobi}, we show the connection between the periods \eqref{eq:T_12}-\eqref{eq:T_34} and the complete elliptic integral of the first kind 
 \begin{equation} 
 {\sf K}(z) \;\equiv\; \int_{0}^{\pi/2}\;\frac{d\phi}{\sqrt{1 \;-\; z\;\sin^{2}\phi}}.
 \label{eq:K_def}
 \end{equation} 
Here, we note that the definition \eqref{eq:K_def} follows the convention used by {\sf Mathematica}. For example, the special lemniscatic case with $g_{3} = 0$ (i.e., $\epsilon = \epsilon_{\delta}$) and $g_{2} = \sin^{2}\varphi = 1/2$ yields the real half-period
 \begin{equation}
 \omega_{0} \;=\; \frac{{\sf K}(1/2)}{\sqrt{\sin\varphi}} \;=\; \frac{\Gamma(\frac{1}{4})^{2}}{2^{7/4}\,\sqrt{\pi}},
 \label{eq:omega0_def}
 \end{equation}
 where $\Gamma(z)$ denotes the gamma function. For the equianharmonic case $g_{2} = 0$ (i.e., $\nu = 0$ when $\epsilon = 1/3$), on the other hand, with $g_{3} = -\,\frac{1}{2}\,\sin^{2}\varphi$ and $\Delta = -\,(27/4)\,\sin^{4}\varphi$, we find the complex half-period
 \begin{equation}
 \omega_{1}^{\rm e} \;=\; \frac{e^{i\pi/12}\;{\sf K}\left(e^{i\pi/3}\right)}{(\sin\varphi)^{\frac{1}{3}}\;\sqrt{\cos(\pi/6)}} = \frac{2^{\frac{1}{6}}\,e^{i\pi/6}\;\Gamma(\frac{1}{3})^{3}}{4\pi\;(\sin\varphi)^{\frac{1}{3}}},
 \end{equation}
 where we used the identity
 \[ {\sf K}\left(e^{i\pi/3}\right) \;\equiv\; e^{i\pi/12}\;\left(\frac{3^{\frac{1}{4}}\, \Gamma\left(\frac{1}{3}\right)^{3}}{2^{\frac{7}{3}}\;\pi}\right), \]
 and the real period for the equianharmonic case is $T_{\rm e} = 2\,{\rm Re}(\omega_{1}^{\rm e})$. The period $T(\epsilon) = 2\,{\rm Re}(\omega_{1})$ is calculated from Eq.~\eqref{eq:omega_1} for $\epsilon > 1/3$, with $\nu = -\,|\nu|$ and $\mu = -\,|\mu|$ (i.e., $g_{2} < 0$ and $g_{3} < 0$), and $e_{1}(\mu,\nu) =  i\,\sqrt{|\nu|/4}\;\cos\left(\frac{1}{3}\,\psi(\eta)\right)$ is calculated from Eq.~\eqref{eq:root_period}.
 
\begin{figure}
\epsfysize=2in
\epsfbox{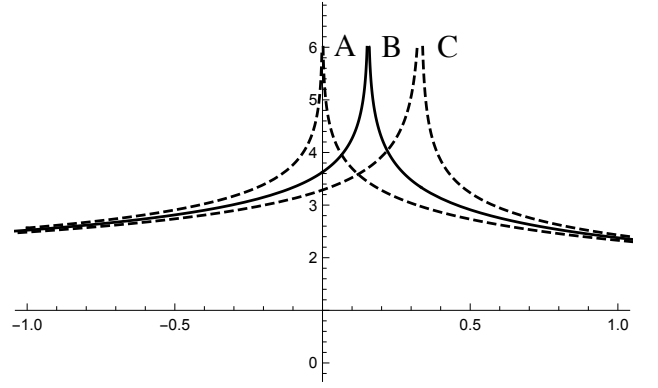}
\caption{Period of oscillation $T(\mu,\nu)$ versus dimensionless energy $\epsilon$ in the range $-1 \leq \epsilon \leq 1$ for $\delta = 0$ (A), $\delta = 1/\sqrt{2}$ (B), and $\delta = 1$ (C). The period 
$T(\mu,\nu) \rightarrow \infty$ as the energy approaches the separatrix value $\epsilon \rightarrow \epsilon_{b}$.}
\label{fig:Periods}
\end{figure}

Figure \ref{fig:Periods} shows the plots of the oscillation period $T(\mu,\nu)$ versus the dimensionless energy $\epsilon$ in the range $-1 \leq \epsilon \leq 1$ for various values of the asymmetry parameter $\delta$. Curve A represents the symmetric case $\delta = 0$, curve B represents the case $\delta = 1/\sqrt{2}$, and curve C represents the case $\delta = 1$ for which the shallow well has just disappeared. The period 
$T(\mu,\nu) \rightarrow \infty$ as the energy approaches the separatrix value $\epsilon \rightarrow \epsilon_{b}$, where $\epsilon_{b} = 0$ (curve A) and $\epsilon_{b} = 1/3$ (curve C).
 
\subsection{Orbits generated from $\xi_{1}$} 

We now consider the Weierstrass solutions for Eq.~\eqref{eq:motion_double_x}. For orbits generated from the turning point $\xi_{1}$, we identify two scenarios: (II.a)-(II.b) for $V_{a} < E < V_{b}$, the second turning point is $\xi_{2}$; (III)-(IV) for $E > V_{b}$, the second turning point is $\xi_{4}$. 

In the energy ranges (II.a)-(II.b), we substitute into Eq.~\eqref{eq:motion_double_x}:
\begin{equation}
x(t) \;\equiv\; \xi_{1} \;+\; \frac{1}{s_{1}(t)}, 
\label{eq:x1_s}
\end{equation}
and we obtain
\begin{eqnarray}
\frac{1}{2}\;\left(\frac{ds_{1}}{dt}\right)^{2} & = & (A_{21}\,s_{1} -1)\;(A_{31}\,s_{1} -1)(A_{41}\,s_{1} -1) \nonumber \\
 & \equiv & a_{3}\,s_{1}^{3} \;+\; a_{2}\;s_{1}^{2} \;+\; a_{1}\;s_{1} \;-\; 1,
\label{eq:motion_double_s1}
\end{eqnarray}
where $A_{k1} \equiv \xi_{k} - \xi_{1}$ and, using the turning-point identities \eqref{eq:turning_identities}, we define the coefficients
\begin{equation}
\left. \begin{array}{rcl}
a_{1} & = & -\,4\;\xi_{1} \;\equiv\; -\;\frac{1}{6}\;V^{\prime\prime\prime}(\xi_{1}) \\
 &  & \\
a_{2} & = & -\,(6\,\xi_{1}^{2} - 3/2) \;\equiv\; -\;\frac{1}{2}\;V^{\prime\prime}(\xi_{1}) \\
 &  & \\
a_{3} & = & -\,4\,\xi_{1}^{3} + 3\,\xi_{1} + \delta \;\equiv\; -\;V^{\prime}(\xi_{1})
\end{array} \right\},
\label{eq:a_123}
\end{equation}
which are expressed in terms of derivatives of the potential \eqref{eq:double_well} evaluated at $\xi_{1}$. 

We transform Eq.~\eqref{eq:motion_double_s1} into the Weierstrass equation 
\begin{equation}
(w^{\prime})^{2} \;=\; 4\,w^{3} \;-\; g_{2}\,w \;-\; g_{3}
\label{eq:w_eq}
\end{equation}
by making the substitution
\begin{equation}
s_{1}(t) \;\equiv\; \frac{1}{a_{3}} \left[ 2\;w(t) \;-\; \frac{a_{2}}{3} \right],
\label{eq:s_wp}
\end{equation}
where the Weierstrass invariants $g_{2}$ and $g_{3}$ are
\begin{equation}
g_{2} \;=\; -\;a_{1}\,a_{3} \;+\; \frac{a_{2}^{2}}{3} \;=\; \frac{3}{4}\;\nu,
\label{eq:g2_double}
\end{equation}
and
\begin{equation}
g_{3} \;=\; \frac{1}{6} \left( 3\,a_{3}^{2} \;-\; \frac{2}{9}\,a_{2}^{3} \;+\; a_{1}a_{2}a_{3} \right) \;=\; \frac{\mu}{8},
\label{eq:g3_double}
\end{equation}
where $\nu$ and $\mu$ are defined in Eq.~\eqref{eq:nu_mu}. The solution for $w(t)$ is expressed in terms of the Weierstrass elliptic function $w(t) = \wp(t ; g_{2}, g_{3})$, which satisfies the initial condition $s_{1}(0) = \infty$ for $x(0) = \xi_{1}$.

The solution for the motion in an asymmetric double-well potential, with initial condition $x(0) = \xi_{1}$, is
\begin{equation}
x(t) \;=\; \xi_{1} \;-\; \frac{V^{\prime}(\xi_{1})}{2\,\wp(t;g_{2},g_{3}) \;+\; V^{\prime\prime}(\xi_{1})/6},
\label{eq:x1_t}
\end{equation}
which is shown in Fig.~\ref{fig:orbits} inside the shallow well (inside the separatrix on the left side) for $\epsilon_{a} < \epsilon < \epsilon_{b}$. The Weierstrass elliptic function satisfies the periodicity condition $\wp(t + T) \equiv \wp(t)$, where the real period is
\begin{eqnarray}
T(\mu,\nu) & = & 2\,\int_{\xi_{1}}^{\xi_{2}}\frac{dx}{\sqrt{2(\xi_{4} - x)(\xi_{3}-x)\,(\xi_{2}-x)(x - \xi_{1})}} \nonumber \\
 & = & \int_{(\xi_{2} - \xi_{1})^{-1}}^{\infty}\frac{\sqrt{2}\;ds}{\sqrt{a_{3} s^{3} + a_{2} s^{2} + a_{1} s - 1}} \label{eq:T_1_def} \\
  & = & 2\int_{e_{1}}^{\infty}\;\frac{dw}{\sqrt{4\,w^{3} - g_{2}\,w - g_{3}}} \equiv 2\;\omega_{1}(g_{2},g_{3}),
  \nonumber 
\end{eqnarray}
which is shown in Table \ref{tab:omega_asymm} and Fig.~\ref{fig:Periods}. In the energy range (II.a), where $(g_{2},g_{3},\Delta) = (+,+,+)$, 
Eq.~\eqref{eq:T_1_def} yields 
\begin{equation}
T_{\rm II.a} \;=\; 2\,\omega_{1}^{+} \;=\; 2\,\omega, 
\label{eq:TW_II.a}
\end{equation}
where $\omega_{1}^{+}$ is defined by Eq.~\eqref{eq:omega_1}. In the energy range (II.b), on the other hand, where $(g_{2},g_{3},\Delta) = (+,-,+)$, 
Eq.~\eqref{eq:T_1_def} yields 
\begin{equation}
T_{\rm II.b} \;=\; 2\,\omega_{1}^{-} \;=\; 2\,(-i\,\omega_{3}^{+}) \;=\; 2\,|\omega^{\prime}|, 
\label{eq:TW_II.b}
\end{equation}
where $\omega_{3}^{+}$ is defined by Eq.~\eqref{eq:omega_3}.

\subsection{Orbits generated from $\xi_{4}$} 

For orbits generated from the turning point $\xi_{4}$, we also identify two scenarios: (I)-(II.a) for $V_{c} < E < V_{a}$, the second turning point is $\xi_{3}$; (II.b)-(IV) for $E > V_{a}$, the second turning point is $\xi_{1}$. If we substitute into Eq.~\eqref{eq:motion_double_x}:
\begin{equation}
x(t) \;\equiv\; \xi_{4} \;-\; \frac{1}{s_{4} (t)}, 
\label{eq:x4_s}
\end{equation}
we obtain
\begin{eqnarray}
\frac{1}{2}\;\left(\frac{ds_{4}}{dt}\right)^{2} & = & (B_{41}\,s_{4} -1)\;(B_{42}\,s_{4} -1)(B_{43}\,s_{4} -1) \nonumber \\
 & \equiv & b_{3}\,s_{4}^{3} \;+\; b_{2}\;s_{4}^{2} \;+\; b_{1}\;s_{4} \;-\; 1,
\label{eq:motion_double_s4}
\end{eqnarray}
where $B_{4k} \equiv \xi_{4} - \xi_{k}$ and, using the turning-point identities \eqref{eq:turning_identities}, we define the coefficients
\begin{equation}
\left. \begin{array}{rcl}
b_{1} & \equiv & 4\;\xi_{4} \;\equiv\; \frac{1}{6}\;V^{\prime\prime\prime}(\xi_{4}) \\
 &  & \\
b_{2} & \equiv & -\,(6\,\xi_{4}^{2} - 3/2) \;\equiv\; -\;\frac{1}{2}\;V^{\prime\prime}(\xi_{4}) \\
 &  & \\
b_{3} & \equiv & 4\,\xi_{4}^{3} - 3\,\xi_{4} - \delta \;\equiv\; V^{\prime}(\xi_{4})
\end{array} \right\},
\label{eq:b_123}
\end{equation}
which are expressed in terms of derivatives of the potential \eqref{eq:double_well} evaluated at $\xi_{4}$. 

We transform Eq.~\eqref{eq:motion_double_s4} into the Weierstrass equation \eqref{eq:w_eq} by making the substitution
\begin{equation}
s_{4}(t) \;\equiv\; \frac{1}{b_{3}} \left[ 2\;\wp(t; g_{2}, g_{3}) \;-\; \frac{b_{2}}{3} \right],
\label{eq:s_wp_4}
\end{equation}
where the Weierstrass invariants $g_{2}$ and $g_{3}$ are
\begin{equation}
g_{2} \;=\; -\;b_{1}\,b_{3} \;+\; \frac{b_{2}^{2}}{3} \;=\; \frac{3}{4}\;\nu,
\label{eq:g2_double_4}
\end{equation}
and
\begin{equation}
g_{3} \;=\; \frac{1}{6} \left( 3\,b_{3}^{2} \;-\; \frac{2}{9}\,b_{2}^{3} \;+\; b_{1}b_{2}b_{3} \right) \;=\; \frac{\mu}{8},
\label{eq:g3_double_4}
\end{equation}
where $\nu$ and $\mu$ are defined in Eq.~\eqref{eq:nu_mu}. Once again, the solution for $w(t)$ is expressed in terms of the Weierstrass elliptic function 
$w(t) = \wp(t ; g_{2}, g_{3})$, which satisfies the initial condition $s_{4}(0) = \infty$ for $x(0) = \xi_{4}$.

The solution for the motion in an asymmetric double-well potential, with initial condition $x(0) = \xi_{4}$, is
\begin{equation}
x(t) \;=\; \xi_{4} \;-\; \frac{V^{\prime}(\xi_{4})}{2\,\wp(t;g_{2},g_{3}) \;+\; V^{\prime\prime}(\xi_{4})/6},
\label{eq:x4_t}
\end{equation}
which is shown in Fig.~\ref{fig:orbits} inside the deep well (inside the separatrix on the right side) for $\epsilon_{c} < \epsilon < \epsilon_{b}$ as well as outside the separatrix for $\epsilon_{b} < \epsilon < 1/3$ and $\epsilon > 1/3$.

The Weierstrass elliptic function satisfies the periodicity condition $\wp(t + T) \equiv \wp(t)$, where the real period is
\begin{eqnarray}
T(\mu,\nu) & = & 2\,\int_{\xi_{3}}^{\xi_{4}}\frac{dx}{\sqrt{2(\xi_{4} - x)(\xi_{3}-x)\,(\xi_{2}-x)(x - \xi_{1})}} \nonumber \\
 & = & \int_{(\xi_{4} - \xi_{3})^{-1}}^{\infty}\frac{\sqrt{2}\;ds}{\sqrt{b_{3} s^{3} + b_{2} s^{2} + b_{1} s - 1}}  \label{eq:T_4_def}  \\
  & = & 2\int_{e_{1}}^{\infty}\;\frac{dw}{\sqrt{4\,w^{3} - g_{2}\,w - g_{3}}} \equiv 2\;\omega_{1}(g_{2},g_{3}).
\nonumber
\end{eqnarray}
In the energy range (I), where $(g_{2},g_{3},\Delta) = (+,+,-)$, Eq.~\eqref{eq:T_4_def} yields 
\begin{equation}
T_{\rm I} \;=\; 2\,\omega_{1}^{+} \;=\; 2\,\Omega. 
\label{eq:TW_I}
\end{equation}
Next, in the energy range (III), where $(g_{2},g_{3},\Delta) = (+,-,-)$, Eq.~\eqref{eq:T_4_def} yields 
\begin{equation}
T_{\rm III} \;=\; 2\,{\rm Re}(\omega_{1}^{-}) \;=\; 2\,{\rm Im}(\omega_{3}^{+}) \;=\; 2\,|\Omega^{\prime}|. 
\label{eq:TW_III}
\end{equation}
We note that both periods \eqref{eq:TW_II.b} and \eqref{eq:TW_III} become infinite as $\epsilon \rightarrow \epsilon_{b}$ from below and above, respectively.

Lastly, in the energy range (IV), where $(g_{2},g_{3},\Delta) = (-,-,-)$, Eq.~\eqref{eq:T_4_def} yields 
\begin{equation}
T_{\rm IV} \;=\; 2\,{\rm Re}(\omega_{1}),
\label{eq:TW_IV}
\end{equation}
where $\omega_{1} = \omega_{3}^{*}$.

\subsection{Single-period motion in an asymmetric double-well potential}

\begin{figure}
\epsfysize=3in
\epsfbox{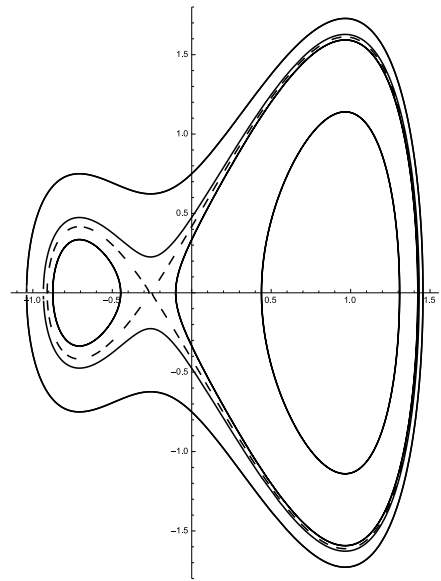}
\caption{Phase plots $(x,\dot{x})$ of the orbit solutions \eqref{eq:x1_t} and \eqref{eq:x4_t} for $\delta = 1/\sqrt{2}$ and $\epsilon_{c} < \epsilon \leq 0.5$, with the separatrix (for $\epsilon = \epsilon_{b}$) shown as a dashed curve. Orbits inside the shallow (deep) well are shown inside the separatrix on the left (right) side.}
\label{fig:orbits}
\end{figure}

The orbit solutions \eqref{eq:x1_t} and \eqref{eq:x4_t} are both expressed in terms of the same Weierstrass elliptic function $\wp(t; g_{2}, g_{3})$ with identical invariants 
$(g_{2},g_{3})$. Figure \ref{fig:orbits} shows the phase plots $(x,\dot{x})$ for these orbit solutions for $\delta = 1/\sqrt{2}$ and $\epsilon$ selected among one of the five energy ranges defined in Eq.~\eqref{eq:epsilon_region}.

\begin{figure}
\epsfysize=2in
\epsfbox{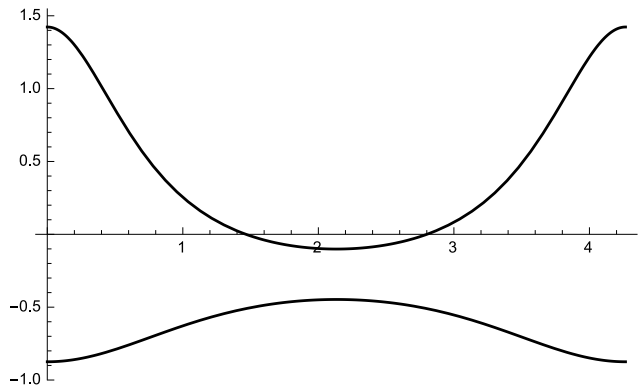}
\caption{Plots of $x(t)$ versus $t$ for $\epsilon_{a} < \epsilon < \epsilon_{b}$: the solution \eqref{eq:x1_t} is shown at the bottom and the solution \eqref{eq:x4_t} is shown at the top.}
\label{fig:monoperiod}
\end{figure}

Figure \ref{fig:monoperiod} shows the orbit solutions \eqref{eq:x1_t} and \eqref{eq:x4_t}  as functions of time $t$ for the case $\epsilon_{a} < \epsilon < \epsilon_{b}$. Here, we clearly see that, while the solutions have different amplitudes ($\xi_{2} - \xi_{1} \neq \xi_{4} - \xi_{3}$), they share the same period 
$T(\epsilon,\delta)$ in agreement with Eq.~\eqref{eq:T_12=34}.

\section{\label{sec:Jacobi}Jacobi Solutions}

\begin{figure}
\epsfysize=2in
\epsfbox{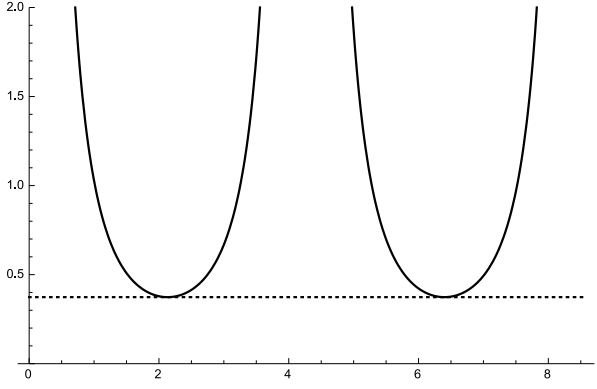}
\caption{Plot of $\wp(t; g_{2},g_{3})$ as a function of time $t$ in the range $0 \leq t \leq 2\,T(\mu,\nu)$. The function periodically returns to its minimum value (shown as a dotted line) at $\wp(\omega_{1}; g_{2},g_{3}) = e_{1} \equiv \chi/2$.}
\label{fig:W_Jacobi}
\end{figure}

The Weierstrass orbit solutions \eqref{eq:x1_t} and \eqref{eq:x4_t} can also be expressed in terms of the Jacobi elliptic functions \cite{Lawden,NIST_Jacobi} as follows. First, we note that $\wp(t; g_{2},g_{3})$ periodically returns to its minimum value (shown as a dotted line in Fig.~\ref{fig:W_Jacobi}) at the cubic root $\wp(\omega_{1}; g_{2},g_{3}) = e_{1}(\mu,\nu)$ defined in Eq.~\eqref{eq:root_period}. The Weierstrass function $\wp(t; g_{2},g_{3})$ can thus be expressed in terms of the Jacobi elliptic function 
${\rm sn}(\kappa t|m)$ as \cite{NIST_Jacobi}
\begin{equation}
\wp(t; g_{2},g_{3}) \;\equiv\; e_{3} \;+\; \frac{e_{1} - e_{3}}{{\rm sn}^{2}(\kappa\,t\;|\;m)},
\label{eq:W_Jacobi}
\end{equation}
where the cubic roots $(e_{1},e_{2},e_{3})$ are defined in Eq.~\eqref{eq:root_period}, with $\kappa(\mu,\nu)$ defined as
\begin{eqnarray}
\kappa^{2}(\mu,\nu) & \equiv & e_{1}(\mu,\nu)  - e_{3}(\mu,\nu) \nonumber \\
 & = & \left(\frac{3}{4}\,\nu\right)^{\frac{1}{2}}\; \sin\left(\frac{\pi}{3} \;+\; \frac{1}{3}\;\psi(\mu,\nu)\right), 
\label{eq:kappa_Jacobi}
\end{eqnarray} 
and the Jacobi modulus $m(\mu,\nu)$ is defined as
\begin{eqnarray}
m(\mu,\nu) & \equiv & \frac{e_{2}(\mu,\nu)  - e_{3}(\mu,\nu)}{e_{1}(\mu,\nu)  - e_{3}(\mu,\nu)} \nonumber \\
 & = & \frac{\sin\left(\frac{1}{3}\psi(\eta)\right)}{\sin\left(\frac{\pi}{3} + \frac{1}{3}\psi(\eta)\right)}.
\label{eq:m_Jacobi}
\end{eqnarray}
where the phase $\psi(\eta)$ defined in Eq.~\eqref{eq:psi_path}. It is also useful to define the complementary modulus
\begin{equation}
m^{\prime}(\mu,\nu) \equiv 1 - m(\mu,\nu) \;=\; \frac{\sin\left(\frac{\pi}{3} - \frac{1}{3}\psi(\eta)\right)}{\sin\left(\frac{\pi}{3} + \frac{1}{3}\psi(\eta)\right)}.
\label{eq:m_prime}
\end{equation}

We now use Eq.~\eqref{eq:W_Jacobi} to connect the period of the Weierstrass elliptic function $\wp(t; g_{2}, g_{3})$ with the period of the Jacobi elliptic function 
${\rm sn}^{2}(\kappa\,t|m)$. Using the fact that the Jacobi elliptic function ${\rm sn}^{2}(z + 2{\sf K}|m) = {\rm sn}^{2}(z|m)$ is a periodic function with period 
$2\,{\sf K}(m)$, the Weierstrass period $T(\mu,\nu)$ is, therefore, expressed as
\begin{equation}
T(\mu,\nu) \;\equiv\; 2\,{\rm Re}\left[\frac{{\sf K}\left(m(\mu,\frac{}{}\nu)\right)}{\kappa(\mu,\nu)}\right].
\label{eq:T_Jacobi}
\end{equation}
For example, since $\psi_{i} = 0$ at $\epsilon = \epsilon_{i}$ ($i = a, c$), we find $m_{i} = 0$ and ${\sf K}(0) = \pi/2$. Next, using $\kappa_{i}^{2} = \frac{3}{4}\,
(4 x_{i}^{2} - 1) = \frac{1}{4}\; V^{\prime\prime}(x_{i})$, we recover the standard result ($i = a, c$)
\begin{equation}
T_{i} \;\equiv\; \frac{2\,{\sf K}(0)}{\kappa_{i}} \;=\; \frac{\pi}{\sqrt{V^{\prime\prime}(x_{i})/4}} \;=\; \frac{2\,\pi}{\sqrt{V^{\prime\prime}(x_{i})}}.
\label{eq:Ti_K}
\end{equation}
When $\psi_{b} = \pi$ is substituted into Eq.~\eqref{eq:m_Jacobi}, on the other hand, we find $m_{b} = \sin(\pi/3)/\sin(2\pi/3) = 1$ so that ${\sf K}(1) = \infty$ and the period becomes infinite as $\epsilon$ approaches the separatrix value $\epsilon = \epsilon_{b}$ (see dashed line in Fig.~\ref{fig:Periods}). For all other values of $\epsilon$ defined in 
Eq.~\eqref{eq:epsilon_region}, we need to compute the functions \eqref{eq:kappa_Jacobi}-\eqref{eq:m_Jacobi} in order to compute the period \eqref{eq:T_Jacobi}. 

\begin{figure}
\epsfysize=2in
\epsfbox{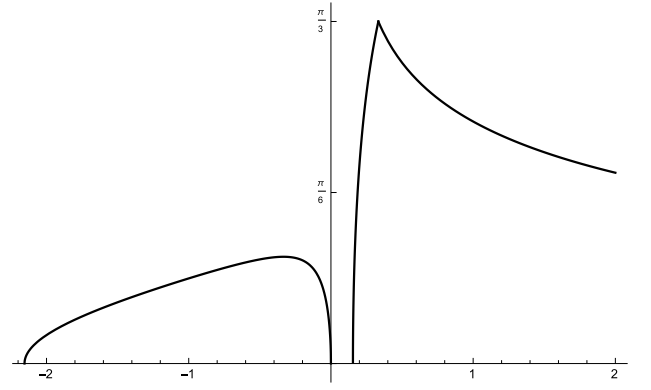}
\caption{Plot of the phase $\theta$ of the Jacobi modulus $m$ for $\epsilon_{c} \leq \epsilon < 2$. The modulus $0 \leq m \leq 1$ is real for $\epsilon_{a} \leq \epsilon \leq \epsilon_{b}$ and, thus, its phase is zero in that range.}
\label{fig:Theta}
\end{figure}

First, in the range (I) $\epsilon_{c} < \epsilon < \epsilon_{a}$, we use $\psi_{I}(\mu,\nu) \equiv i\,\phi_{I}(\mu,\nu)$, where $\phi_{I}(\mu,\nu) \equiv \cosh^{-1}(\mu/\nu^{3/2})$, and we find the complex Jacobi modulus
\begin{equation}
m_{I}(\mu,\nu)  \;\equiv\; 1 \;-\; \exp\left(-i\frac{}{} \theta_{I}(\mu,\nu)\right),
\label{eq:m_I}
\end{equation}
where the real phase $\theta_{I}(\mu,\nu)$ is defined as
\begin{equation}
\theta_{I}(\mu,\nu) \equiv 2\;\tan^{-1}\left[3^{-\frac{1}{2}}\;\tanh\left(\frac{1}{3}\,\phi_{I}(\mu,\nu)\right) \right].
\label{eq:zeta_I}
\end{equation} 
We also find
\begin{eqnarray}
\kappa_{I}^{2}(\mu,\nu) & = & \left(\frac{3}{4}\,\nu\right)^{\frac{1}{2}}\; \sin\left(\frac{\pi}{3} \;+\; \frac{i}{3}\;\phi_{I}(\mu,\nu)\right) 
\label{eq:kappa_I} \\
 & \equiv & |\kappa_{I}|^{2}\; e^{i\theta_{I}/2},
\nonumber
\end{eqnarray}
and the period $T_{I}(\mu,\nu)$ is defined as
\begin{equation}
T_{I}(\mu,\nu) \;=\; \frac{2}{|\kappa_{I}|}\;e^{-i\,\theta_{I}/4}\;{\sf K}\left(1 - e^{-i\,\theta_{I}}\right),
\label{eq:T_I}
\end{equation}
where $e^{-i\,\theta_{I}/4}\;{\sf K}\left(1 - e^{-i\,\theta_{I}}\right)$ is real for $0 \leq \theta_{I} \leq \pi$. Figure \ref{fig:Theta} shows that $\theta_{I} < \pi/9$, for 
$\delta = 1/\sqrt{2}$ and $\epsilon_{c} < \epsilon < \epsilon_{a}$, and, hence, the period \eqref{eq:T_I} is real.

Next, in the ranges (II.a) $\epsilon_{a} \leq \epsilon \leq \epsilon_{\delta}$ and (II.b) $\epsilon_{\delta} \leq \epsilon \leq \epsilon_{b}$, the phase 
$\psi_{II}(\mu,\nu) = \cos^{-1}(\mu/\nu^{3/2})$ is real and
\begin{equation}
0 \;\leq\; m_{II}(\mu,\nu) \;=\; \frac{\sin\left(\frac{1}{3}\psi_{II}(\mu,\nu)\right)}{\sin\left(\frac{\pi}{3} + \frac{1}{3}\psi_{II}(\mu,\nu)\right)} \;\leq\; 1,
\end{equation}
while
\begin{equation}
\kappa_{II}^{2}(\mu,\nu) \;=\; \left(\frac{3}{4}\,\nu\right)^{\frac{1}{2}}\; \sin\left(\frac{\pi}{3} \;+\; \frac{1}{3}\;\psi_{II}(\mu,\nu)\right).
\end{equation}
In these ranges, the period $T_{II}(\mu,\nu)$ is, therefore, expressed as
\begin{equation}
T_{II}(\mu,\nu) \;=\; 2\,{\sf K}(m_{II})/\kappa_{II}.
\label{eq:T_II}
\end{equation}
For the special lemniscatic case $\epsilon = \epsilon_{\delta}$ (i.e., $g_{3} = 0 = \mu_{\delta}$), for which $\psi_{\delta} = \pi/2$ and $\nu_{\delta} = (4/3)\,\sin^{2}\varphi$, we find $m_{\delta} = 1/2$, $\kappa_{II}^{2} = \sin\varphi$,  and $T_{\delta} = 2\,{\sf K}(1/2)/(\sin\varphi)^{\frac{1}{2}}$, in agreement with Eq.~\eqref{eq:omega0_def}.

For the range (III) $\epsilon_{b} \leq \epsilon < 1/3$, the phase $\psi_{III}(\mu,\nu) \equiv \pi - i\,\phi_{III}(\mu,\nu)$ is complex-valued, where $\phi_{III}(\mu,\nu) \equiv \cosh^{-1}(|\mu|/\nu^{3/2})$, and the complex-valued Jacobi modulus is
\begin{equation}
m_{III}(\mu,\nu)  \;\equiv\; \exp\left(-i\frac{}{} \theta_{III}(\mu,\nu)\right),
\label{eq:m_III}
\end{equation}
where the real phase $0 \leq \theta_{III}(\mu,\nu) < \pi/3$ is defined as
\begin{equation}
\theta_{III}(\mu,\nu) \equiv 2\;\tan^{-1}\left[3^{-\frac{1}{2}}\;\tanh\left(\frac{1}{3}\,\phi_{III}(\mu,\nu)\right) \right].
\label{eq:zeta_III}
\end{equation} 
We also find
\begin{eqnarray}
\kappa_{III}^{2}(\mu,\nu) & = & \left(\frac{3}{4}\,\nu\right)^{\frac{1}{2}}\; \sin\left(\frac{2\pi}{3} \;-\; \frac{i}{3}\;\phi_{III}(\mu,\nu)\right) 
\nonumber \\
 & \equiv &  |\kappa_{III}|^{2}\; e^{i\theta_{III}/2},
\label{eq:kappa_III} 
\end{eqnarray}
and the period \eqref{eq:T_Jacobi} is now defined as
\begin{equation}
T_{III}(\mu,\nu) \;=\; \frac{2}{|\kappa_{III}|}\;{\rm Re}\left[ e^{-i\,\theta_{III}/4}\;{\sf K}\left(e^{-i\,\theta_{III}}\right) \right].
\label{eq:T_III}
\end{equation}
In the equianharmonic limit $\epsilon \rightarrow 1/3$ (i.e., $g_{2} = 0$), we find $\theta_{III} \rightarrow \pi/3$ (see Fig.~\ref{fig:Theta}), and $|\kappa_{III}| \rightarrow 3^{\frac{1}{4}}/2^{\frac{2}{3}}$.

Lastly, in the range (IV) $\epsilon > 1/3$,  the phase $\psi_{IV}(\mu,\nu) \equiv \pi/2 + i\,\phi_{IV}(\mu,\nu)$ is complex-valued, where $\phi_{IV}(\mu,\nu) \equiv \sinh^{-1}(|\mu|/
|\nu|^{3/2})$, and the complex-valued Jacobi modulus is
\begin{eqnarray}
m_{IV}(\mu,\nu)  & \equiv & \frac{1}{2} \;+\; i\;\frac{\sqrt{3}}{2}\;\tanh\left(\frac{1}{3}\,\phi_{IV}(\mu,\nu) \right) \nonumber \\
 & = & |m_{IV}|\;e^{i\,\theta_{IV}},
\label{eq:m_IV}
\end{eqnarray}
where
\begin{equation}
\left. \begin{array}{rcl}
|m_{IV}|(\mu,\nu) & = & \frac{1}{2} \left[ 1 \,+\, 3\;\tanh^{2}\left(\frac{1}{3}\,\phi_{IV}(\mu,\nu) \right) \right]^{1/2} \\
 &  & \\
\theta_{IV}(\mu,\nu) & = & \tan^{-1}\left[ \sqrt{3}\;\tanh\left(\frac{1}{3}\,\phi_{IV}(\mu,\nu) \right)\right]
\end{array} \right\},
\end{equation}
and
\begin{eqnarray}
\kappa_{IV}^{2}(\mu,\nu) & = & i\;\left(\frac{3}{4}\,|\nu|\right)^{\frac{1}{2}}\; \sin\left(\frac{\pi}{2} \;+\; \frac{i}{3}\;\phi_{IV}(\mu,\nu)\right) 
\nonumber \\
 & \equiv & i\;\left(\frac{3}{4}\,|\nu|\right)^{\frac{1}{2}}\; \cosh\left(\frac{1}{3}\;\phi_{IV}(\mu,\nu)\right).
\label{eq:kappa_IV}
\end{eqnarray}
The period \eqref{eq:T_Jacobi} is, therefore, defined as
\begin{equation}
T_{IV}(\mu,\nu) \;=\; \frac{2}{|\kappa_{IV}|}\;{\rm Re}\left[ e^{-i\pi/4}\;{\sf K}\left(m_{IV}\right) \right].
\label{eq:T_IV}
\end{equation}
Hence, the periods shown in Table \ref{tab:omega_asymm} (and in Fig.~\ref{fig:Periods}) can all be expressed in terms of the complete elliptic integral of the first kind \eqref{eq:K_def} according to the relation \eqref{eq:T_Jacobi}.

\section{\label{sec:sum}Summary}

In the present paper, we have presented a complete solution for the periodic orbits of a particle moving in an asymmetric double-well potential represented by 
Eq.~\eqref{eq:double_well}. In Secs.~\ref{sec:min_max} and \ref{sec:energy}, we first derived explicit expressions for the extrema \eqref{eq:V_prime_roots} of the asymmetric double-well potential \eqref{eq:double_well} and then we derived explicit expressions for the four turning points \eqref{eq:xi_1}-\eqref{eq:xi_4} as functions of the normalized energy $\epsilon = 16\,E/9$ and the asymmetry parameter $|\delta| < 1$. In Sec.~\ref{sec:Weierstrass}, the solutions \eqref{eq:x1_t} and \eqref{eq:x4_t} were expressed in terms of the unique 
Weierstrass elliptic function $\wp(t; g_{2},g_{3})$, where the Weierstrass invariants $g_{2} = 3\,(1 - 3 \epsilon)/4$ and $g_{3} = (4\,\delta^{2} - 1 - 9\epsilon)/8$ depend on the orbit parameters $(\epsilon,\delta)$.

When the four turning points \eqref{eq:xi_1}-\eqref{eq:xi_4} are real, the periodic motions in the deep and shallow wells of the asymmetric double-well potential were shown to have identical periods $T_{12} = T_{34}$ at the same energy level. This result was demonstrated in Eqs.~\eqref{eq:T_1_def} and \eqref{eq:T_4_def}, which is consistent with the equivalency \eqref{eq:T_12=34} established by the continuous deformation of contours $C_{12} \leftrightarrow C_{34}$ on the Riemann sphere. Hence, this period equivalency might be used as an argument for the physical reality of the Riemann sphere. Using this equivalency and the residue theorem from complex analysis, the standard result $T_{0} = 2\pi/\sqrt{V^{\prime\prime}(x_{0})}$ was re-derived for the shallow and deep wells in Eqs.~\eqref{eq:T_a} and \eqref{eq:T_c}, respectively.

In Sec.~\ref{sec:Jacobi}, we used the standard connection \eqref{eq:W_Jacobi} between the Weierstrass and Jacobi elliptic functions to obtain an explicit expression for the period $T(\epsilon,\delta)$ in terms of the complete elliptic integral of the first kind, ${\sf K}(m)$, where the Jacobi modulus $m(\epsilon,\delta)$ depends on the orbit parameters $(\epsilon,\delta)$.

Lastly, we briefly discuss an application of the present work by considering the nonlinear undamped Duffing equation \cite{Holmes_1979,LL_chaos,Litak} driven by a time-dependent asymmetry perturbation:\begin{equation}
\ddot{x} \;-\; 3\,x \;+\; 4\,x^{3} \;=\; \delta(t),
\label{eq:x_ddot}
\end{equation}
which is derived from the double-well potential \eqref{eq:double_well} as an Euler-Lagrange equation. Here, when $\delta(t)$ is a periodic function of time (e.g., $\delta = \delta_{0}\,\cos\omega_{0}t$, where 
$\delta_{0}$ and $\omega_{0}$ are constants), chaotic behavior can be investigated with the Melnikov method \cite{Holmes_1979,LL_chaos}. One possible extension of this classical work involves an asymmetry perturbation represented by an elliptic function $\delta(t) = \delta_{0}\,{\rm cn}(\omega_{0}t|m_{0})$, where we recover the previous work  \cite{Holmes_1979,LL_chaos} when the Jacobi modulus $m_{0} = 0$.

\appendix

\section{\label{sec:App_A}Weierstrass Invariants and Roots}

The solution of a cubic equation is greatly facilitated by the trigonometric identity
\begin{equation}
\cos\phi \;=\; 4\;\cos^{3}(\phi/3) \;-\; 3\;\cos(\phi/3),
\end{equation}
which implies that $x_{0} = \cos(\phi/3)$ is one solution of the cubic equation $4\,x^{3} - 3\,x - \cos\phi = 0$. The remaining two roots are $x_{\pm} = -\,\cos[(\pi \pm \phi)/3]$.

The cubic equation \eqref{eq:cubic_eq} is a special case of the Weierstrass cubic equation
\begin{equation}
4\;x^{3} \;-\; g_{2}\;x \;-\; g_{3} \;=\; 0,
\label{eq:W_cubic}
\end{equation}
where the invariants $g_{2} \equiv 3\,\beta^{2}$ and $g_{3} \equiv \beta^{3}\,\cos\phi$ are defined in terms of two parameters $(\beta,\phi)$, whose roots $(e_{1},e_{2},e_{3})$ are expressed as
\begin{equation}
\left. \begin{array}{rcl}
e_{1} & = & \beta\;\cos(\phi/3) \\
e_{2} & = & -\,\beta\;\cos[(\pi + \phi)/3] \\
e_{3} & = & -\;\beta\;\cos[(\pi - \phi)/3]
\end{array} \right\},
\label{eq:W_roots}
\end{equation}
so that $e_{1} + e_{2} + e_{3} = 0$ and the discriminant is
\begin{equation}
\Delta \;\equiv\; g_{2}^{3} - 27\,g_{3}^{2} \;=\; 27\,\beta^{6}\,\sin^{2}\phi. 
\label{eq:Delta_general}
\end{equation}

For a general cubic equation of the form 
\begin{equation}
P(y) \;\equiv\; 4\,y^{3} \;+\; a\,y^{2} \;+\; b\,y + c, 
\label{eq:cubic_generic}
\end{equation}
where $(a,b,c)$ are arbitrary coefficients, we can use the transformation $y = x + \alpha$, with $\alpha = -\,a/12$ defined by $P^{\prime\prime}(\alpha) \equiv 0$, to obtain the Weierstrass cubic equation \eqref{eq:W_cubic}, with $g_{2} = -\,P^{\prime}(\alpha)$ and $g_{3} = -\,P(\alpha)$. Hence, the roots of Eq.~\eqref{eq:cubic_generic} are $y_{i} = e_{i} + \alpha$ $(i = 1, 2, 3)$, where the Weierstrass roots $e_{i}$ are given in Eq.~\eqref{eq:W_roots}, with $\beta \equiv \sqrt{g_{2}/3}$ and $\cos\phi = g_{3}/\beta^{3}$.

\section{\label{sec:App_B}Symmetric Case}

In the symmetric case $(\delta = 0)$, the four roots of the energy equation 
\begin{equation}
\xi_{k0}^{4} \;-\; \frac{3}{2}\,\xi_{k0}^{2} \;-\; \frac{9\,\epsilon}{16} \;=\; 0.
\label{eq:energy_quartic_B}
\end{equation}
are $\pm\;\frac{1}{2}\,\sqrt{3}\;( 1 \;\pm\; \sqrt{1 + \epsilon})^{1/2}$, which are all real if $-1 < \epsilon < 0$. In this range, we use $\epsilon \equiv -\,\sin^{2}\alpha$ (with $0 <\alpha < \pi/2$), so that the four roots become
\begin{eqnarray}
\xi_{10} & = & -\;\sqrt{\frac{3}{2}}\;\cos\frac{\alpha}{2}, \label{eq:xi_10_alpha} \\
\xi_{20} & = & -\;\sqrt{\frac{3}{2}}\;\sin\frac{\alpha}{2}, \label{eq:xi_20_alpha} \\
\xi_{30} & = & \sqrt{\frac{3}{2}}\;\sin\frac{\alpha}{2}, \label{eq:xi_30_alpha} \\
\xi_{40} & = & \sqrt{\frac{3}{2}}\;\cos\frac{\alpha}{2}, \label{eq:xi_40_alpha}
\end{eqnarray}
where $\sin\alpha/2 < \cos\alpha/2$ in the range $0 <\alpha < \pi/2$. In the case $\epsilon > 0$, only the two roots 
$\pm\;\frac{1}{2}\,\sqrt{3}\;( 1 + \sqrt{1 + \epsilon})^{1/2}$ are real, while the other two roots $\pm\;\frac{1}{2}\,\sqrt{3}\;( 1 - \sqrt{1 + \epsilon})^{1/2}$ are purely imaginary.

We now consider the symmetric limit of our general roots \eqref{eq:xi_1}-\eqref{eq:xi_4} for the case $\epsilon = -\,\sin^{2}\alpha$. First, we use the cubic root of Eq.~\eqref{eq:chi_cubic}:
\begin{equation}
\chi \;=\; \frac{1}{2} \;+\; \frac{3}{2}\;\sin\alpha \;\equiv\; 4\,\sigma^{2} \;-\; 1 ,
\label{eq:chi_alpha}
\end{equation}
which yields
\begin{equation}
\left. \begin{array}{rcl}
\sigma & = & \sqrt{(3/8)\;(1 + \sin\alpha)} \\
 &  & \\
\frac{1}{2}\,\sqrt{3 - 4\,\sigma^{2}} & = & \sqrt{(3/8)\;(1 - \sin\alpha)}
\end{array}\right\},
\label{eq:sigma_alpha}
\end{equation}
and Eqs.~\eqref{eq:xi_1}-\eqref{eq:xi_4} yield
\begin{eqnarray}
\xi_{10} & = & -\;\sqrt{\frac{3}{8}}\;\left( \sqrt{1 + \sin\alpha} \;+\frac{}{} \sqrt{1 - \sin\alpha}\right), \label{eq:xi10_alpha} \\
\xi_{20} & = & -\;\sqrt{\frac{3}{8}}\;\left( \sqrt{1 + \sin\alpha} \;-\frac{}{} \sqrt{1 - \sin\alpha}\right), \label{eq:xi20_alpha} \\
\xi_{30} & = & \sqrt{\frac{3}{8}}\;\left( \sqrt{1 + \sin\alpha} \;-\frac{}{} \sqrt{1 - \sin\alpha}\right), \label{eq:xi30_alpha} \\
\xi_{40} & = & \sqrt{\frac{3}{8}}\;\left( \sqrt{1 + \sin\alpha} \;+\frac{}{} \sqrt{1 - \sin\alpha}\right). \label{eq:xi40_alpha}
\end{eqnarray}
Lastly, by using the identities $\sqrt{1 \pm \sin\alpha} \equiv \cos(\alpha/2) \pm \sin(\alpha/2)$, we recover the symmetric roots \eqref{eq:xi_10_alpha}-\eqref{eq:xi_40_alpha} from Eqs.~\eqref{eq:xi10_alpha}-\eqref{eq:xi40_alpha}.

\section{\label{sec:quartic}Motion in a Symmetric Quartic Potential}

We look at particle orbits in the (dimensionless) quartic potential $V(x) = x^{4} - 3\,x^{2}/2$. Here, the turning points for $E \equiv 
9\,\epsilon/16 = V(x)$ are
\begin{equation}
\left. \begin{array}{lr}
\pm\,\frac{1}{2}\;\sqrt{3\,(1 + {\sf e})} & ({\rm for}\;{\sf e} > 1) \\
 & \\
0 \;{\rm and}\; \pm\,\sqrt{3/2} & ({\rm for}\;{\sf e} = 1) \\
 & \\
\pm\,\frac{1}{2}\;\sqrt{3\,(1 \pm {\sf e})} & ({\rm for}\;{\sf e} < 1)
\end{array} \right\},
\label{eq:tp_quartic}
\end{equation}
where ${\sf e} \equiv \sqrt{1 + \epsilon}$. Each orbit is solved using the initial conditions $x_{0} = \frac{1}{2}\,\sqrt{3\,(1 + {\sf e})} \equiv a$ and $\dot{x}_{0} = 0$.

\subsection{Jacobi elliptic solution}

The solution for the orbit in a symmetric double-well potential can be given in terms of Jacobi elliptic functions as follows. First, we consider the integral solution
\begin{eqnarray}
t(x) & = & -\;\frac{1}{2}\;\int_{\frac{1}{2}\sqrt{3\,(1 + {\sf e})}}^{x}\;\frac{dy}{\sqrt{\frac{9}{16}({\sf e}^{2} - 1) \;+\; y^{2}\,
(\frac{3}{2} - y^{2})}} \nonumber \\
 & = & \frac{1}{\sqrt{3\,{\sf e}}}\;\int_{0}^{\Phi(x)}\; \frac{d\varphi}{\sqrt{1 \;-\; m\;\sin^{2}\varphi}},
\label{eq:quartic_first}
\end{eqnarray}
where $m \equiv (1 + {\sf e})/2{\sf e}$ while we used the trigonometric substitution $y = \frac{1}{2}\,\sqrt{3\,(1 + {\sf e})}\,\cos\varphi$ with
\begin{equation}
\Phi(x) \;\equiv\; \cos^{-1}\left[\frac{2\;x}{\sqrt{3\,(1 + {\sf e})}}\right] 
\label{eq:Phi4_def}
\end{equation}
to obtain the last expression in Eq.~(\ref{eq:quartic_first}). 

For ${\sf e} > 1$ (i.e., $\epsilon > 0$ and $m < 1$), we find 
\[ \sin\Phi(x) \;=\; {\rm sn}(\sqrt{3\,{\sf e}}\,t|m) \;=\; \sqrt{1 \;-\; \frac{4\;x^{2}(t)}{3\,(1 + {\sf e})}}, \]
which yields 
\begin{equation}
x(t) \;=\; \frac{1}{2}\,\sqrt{3\,(1 + {\sf e})}\;\;{\rm cn}(\sqrt{3\,{\sf e}}\,t|m).
\label{eq:quartic_large}
\end{equation}
For ${\sf e} = 1$ (i.e., the separatrix orbit with $\epsilon = 0$ and $m = 1$), we find
\begin{equation}
x(t) \;=\; \sqrt{3/2}\;\;{\rm sech}(\sqrt{3}\;t).
\label{eq:quartic_sep}
\end{equation}
Lastly, for ${\sf e} < 1$ (i.e., $-1 < \epsilon < 0$ and $m > 1$), we find
\begin{equation}
x(t) \;=\; \frac{1}{2}\,\sqrt{3\,(1 + {\sf e})}\;\;{\rm dn}\left(\sqrt{(3/2)\,({\sf e} + 1)}\;t\,|m^{-1}\right).
\label{eq:quartic_small}
\end{equation}

Except for the separatrix case ${\sf e} = 1$, the motion is periodic, with periods
\begin{equation}
T({\sf e}) \;=\; \left\{ \begin{array}{lr}
4\,{\sf K}(m)/\sqrt{3\,{\sf e}} & \;\;\;({\sf e} > 1) \\
 & \\
 2\,{\sf K}(m^{-1})/\sqrt{3\,{\sf e}\,m} & \;\;\;\;({\sf e} < 1)
 \end{array} \right. 
 \end{equation}
 expressed in terms of the complete elliptic integral of the first kind ${\sf K}(m)$, with ${\sf K}(0) = \pi/2$ and $\lim_{m\rightarrow 1}{\sf K}(m) = \infty$. We note that the period of the orbit corresponding to $\epsilon = -1$ (at the bottom of the well), with ${\sf e} = 0 = m^{-1}$ and ${\sf e}\,m = 1/2$, is $T(-1) = 
 \pi/\sqrt{3/2} = 2\pi/\sqrt{6} \neq 0$. This result agrees with the standard result $T = 2\pi/\sqrt{V^{\prime\prime}(a)}$, where $V^{\prime\prime}(a) = 12\,a^{2} - 3 = 6$.
 
\subsection{Weierstrass elliptic solution}

The Weierstrass solution of the symmetric double-well problem is simply expressed as
\begin{equation}
x(t) \;=\; a \;-\; \frac{3\,a\,{\sf e}}{2\,\wp(t; g_{2}, g_{30}) + (1 + 3\,{\sf e}/2)}, 
\label{eq:wp_symm}
\end{equation}
where  $g_{2} = 3\,\nu/4 = 3\,(1 - 3\,\epsilon)/4$ and $g_{30} = \mu_{0}/8 = -\,(1 + 9\,\epsilon)/8$, with modular discriminant $\Delta_{0} = g_{2}^{3} - 27\,g_{30}^{2} = - (27/8)^{2}\epsilon\,(1 + \epsilon)^{2}$. At the bottom of the double-well potential, where $\epsilon = -1$ and ${\sf e} = 0$, with $(g_{2},g_{30},\Delta_{0}) = (3,1,0)$, the solution \eqref{eq:wp_symm} yields $x(t) = a$ for all times, while the limiting period is $T = 
2\,\omega_{1}(3,1) = 2\pi/\sqrt{6}$.

\end{document}